\documentclass[10pt]{article}
\usepackage{epsfig}
\begin{document}

\twocolumn[\hsize\textwidth\columnwidth\hsize\csname@twocolumnfalse\endcsname

\begin{center}

{ \bf\large Kerr-lens mode-locked lasers as transfer oscillators for optical frequency measurements}
\vspace{0.5cm}

Harald R. Telle, Burghard Lipphardt and J\"orn Stenger
\vspace{0.25cm}

{\it Physikalisch-Technische Bundesanstalt, Bundesallee 100, 38116 Braunschweig, Germany}
\vspace{1cm}

\begin{minipage}{12cm}
We introduce a novel concept for optical frequency measurement and division which employs a 
Kerr-lens mode-locked laser as a transfer oscillator whose noise properties do not enter the measurement 
process. We experimentally demonstrate, that this method opens up the route to phase-link signals with 
arbitrary frequencies in the optical or microwave range while their frequency stability is preserved.  \\

PACS numbers: 42.62.Eh, 06.20.Fn, 06.30.Ft
\end{minipage}
\vspace{1cm}

\end{center}

]

\noindent {\bf 1. Introduction }
 
Continuous-wave  mode-locked lasers emit a periodic train of short pulses. The spectrum of this emission 
corresponds to a comb of distinct lines with well-defined spacing if the mode-coupling mechanism is 
sufficiently strong and fast.  It has been 
shown that the non-resonant, electronic Kerr-lens mode-locking (KLM) process can satisfy this requirement
[\ref{udem99}], making such lasers highly suited for optical frequency measurements [\ref{udem01}], 
[\ref{sten01b}], [\ref{nier00}]. 
The frequency of any line of the spectral comb emitted by such a KLM laser is given by an integer order 
number $m$, the pulse repetition frequency $f_{rep}$ and a so called carrier-envelope offset-frequency 
$\nu_{ceo}$, which accounts for the offset of the entire comb with respect to the frequency zero:

\begin{equation}
\nu_m = \nu_{ceo} + m f_{rep} \,.	
\label{num}
\end{equation}	

When all three quantities are known, any unknown external optical frequency $\nu_x$ within the span 
of the comb can be absolutely measured by detection of its beat-note frequency $\Delta_x$ with a 
suitable comb line.

Absolute frequency measurement means that $f_{rep}$ is measured and expressed in terms of SI-Hertz. 
This primary unit is realized by a Cs-clock controlled radio frequency (rf) source, which is a H-maser 
generating a standard frequency of 100 MHz in our case. 
Thus, no frequency measurement can be better in fractional frequency instability 
than that of the rf-reference. However, optical frequencies are measured in many cases with respect 
to an optical reference frequency by measurement of their frequency difference [\ref{udem99}]. 
Then, only a fraction of the H-maser noise enters the 
measurement process, given by the ratio between the frequency difference and the absolute frequency. 

The limitation due to the H-maser noise can be overcome by measurement of frequency {\it ratios} 
rather than frequency {\it differences} whenever oscillators with better noise properties than the 
H-maser are to be compared. 
A frequency ratio is unitless, i.e. there is no need to refer to the unit Hertz, and thus the frequency 
noise properties of the oscillators involved can be preserved when building the ratio. 
We will demonstrate below, that, in fact, optical 
frequency ratios can be measured with much smaller instabilities than that of the H-maser.

An important issue for such measurements is the frequency noise of the KLM laser due to technical perturbations. 
Conventional approaches [\ref{jone00}] attempt to stabilize both the group and phase delay of the laser resonator 
by piezo transducers. 
However, as a result of the finite response time of these elements, the servo bandwidth of such 
servo loops is typically not sufficient to reduce the frequency noise of the beat-note $\Delta_x$ to a 
level below the noise of the optical signal at $\nu_x$. 
For the same reason, it is very difficult to reduce the fractional frequency noise of $f_{rep}$ to levels 
below that of the microwave reference, at least at high Fourier frequencies. 
As a consequence, the short-term instability of those measurements is limited by the mode-locked laser 
and long averaging times are required for measurements with low uncertainties. 

Our novel approach completely differs in handling the technical frequency fluctuations of the KLM laser. 
We generalize the 
transfer oscillator concept [\ref{kram92}], which relates signals with integer frequency ratios, 
to signals with rational frequency ratios. 
Here, the laser is only slowly frequency stabilized while all beat-notes are phase-tracked with fast 
phase-locked loops (PLL) and online processed with analog electronics. 
Thus, we are not any more limited by the inertia of the mirror 
transducers but can make use of the large signal-to-noise ratios of our beat notes which allow for wide 
servo bandwidths and thus small residual error signals. 
As a result, the additive noise of the measurement process becomes substantially 
smaller than the frequency noise of the signals involved and can be neglected. 
This novel concept, which compensates the noise of the mode-locked laser will be named transfer concept 
in the following. It will be applied to the measurement of frequency ratios between various 
frequency standards: a diode laser at 871~nm, a Nd:YAG laser at 1064~nm, a dye laser operating at 657~nm 
and a 100~MHz reference signal from a H-maser.

\vspace{3ex}
\noindent {\bf 2. Elastic tape picture and transfer concept }
\vspace{1ex}

We integrate Eqn.~(\ref{num}) to relate the instantaneous phases of all signals, 

\begin{equation}
\varphi_m(t) = \varphi_{ceo}(t) + m  \varphi_{rep}(t) + \phi(m),
\label{phim}
\end{equation}

\noindent where $\varphi_m$, $\varphi_{ceo}$, and $\varphi_{rep}$ denote the instantaneous phase angles 
of $\nu_m$, $\nu_{ceo}$, and $\nu_{rep}$, respectively. 
The integration constant $\phi(m)$ accounts for the dispersion properties of the optical components involved. 
It is assumed to be constant or only slowly time-varying. 
This ansatz is motivated by the fact, that the fast electronic Kerr effect in KLM lasers tightly couples 
almost instantaneously all modes of the comb. 
In other words, any individual mode is injection-locked by a strong input signal resulting from the 
superimposed modulation side-bands of the other modes. 
Thus, one expects that the quantum-limited carrier frequency noise is determined by almost the total laser power, 
similar to the Schawlow-Townes limit of a single frequency laser. 
In this sense, the KLM laser oscillation can be considered as one spectrally extended super-mode.
Given the validity of Eqn.~(\ref{phim}), frequency fluctuations of this super-mode resulting from technical 
perturbations can be expressed by various pairs of orthogonal components, e.g. 
\begin{enumerate}
\item[{\bf A.}] common mode fluctuations, i.e. fluctuations of the mean of the group- and phase delay of the laser 
resonator while the ratio of both quantities remains constant and
\item[{\bf B.}] fluctuations of the difference of these quantities while one of them, e.g. the phase delay, 
is held constant. 
\end{enumerate}

This behaviour can be visualized as an elastic tape labelled with a scale of equidistant spectral lines 
which is randomly stretched while it is held fixed at a characteristic point $\nu_{fix}$ on the frequency scale. 
This fixed frequency characterizes the specific type of technical noise. 
In case A $\nu_{fix}=0$ whereas $\nu_{fix}$ is found in the optical carrier frequency region, 
$\nu_{fix}=\nu_{car}$, in case B. 
For acoustic vibrations of resonator mirrors, as an example, we find a fractional change of the 
phase delay $\tau_p$:

\begin{equation}
\left| \frac{\Delta\tau_p}{\tau_p}\right| = \left| \frac{\Delta\nu_{car}}{\nu_{car}}\right|  
           = \left| \frac{\Delta z}{z+n_pz_m}\right| \;,
\label{fixakustikp}
\end{equation}

\noindent
where $z$ is the cavity length, $z_m$ the length of the gain medium and $n_p$ its phase-index of refraction. 
The refractive index of the air and possible other intra-cavity elements have been neglected 
in Eqn.~(\ref{fixakustikp}).
The corresponding expression for the fractional change of the group delay $\tau_g$ reads

\begin{equation}
\left| \frac{\Delta\tau_g}{\tau_g}\right| = \left| \frac{\Delta f_{rep}}{f_{rep}}\right|  
           = \left| \frac{\Delta z}{z+n_gz_m}\right| \;.
\label{fixakustikg}
\end{equation}

\noindent
The fixed point frequency is found

\begin{equation}
\nu_{fix} = \nu_{car} \left( 1-\frac{\Delta\nu_{car}/\nu_{car}}{\Delta f_{rep}/f_{rep}} \right)
            \approx  \nu_{car} (n_p-n_g) \frac{z_m}{z} \;.
\label{nufixakustik}
\end{equation}

\noindent
For the parameters of our laser, we obtain 
\begin{equation}
\nu_{fix}  \approx  - 50 \;{\rm GHz} \qquad\mbox{ (mirror vibration)}.
\end{equation}

\noindent This frequency is small compared to the carrier frequency of a few hundred THz. 
Thus, cavity length fluctuations represent case A to good approximation. 
As a consequence the comb offset frequency $\nu_{ceo}$, which is by definition a frequency close to zero, 
is only weakly affected by such fluctuations.
Fixed point frequencies of other types of fluctuations can be estimated in a similar manner. 
For example, temperature changes of the gain medium lead to a change of its physical length, 
a change of the phase index and a different change of the group index. We estimate a fixed point frequency of
\begin{eqnarray}
&&\nu_{fix}   \approx  20 \;{\rm THz.} \\ 
               & & \mbox{ (temperature variation of gain medium)} \nonumber
\end{eqnarray}

\noindent 
for the parameters of our laser. 

As another specific perturbation, tilting of the cavity mirror behind the double-prism arrangement [\ref{reic99}], 
is of particular interest. 
The cavity length is not affected for a specific carrier frequency but owing to the lateral spectral spread of the 
laser mode at this position, the group delay is strongly changed by such tilting. 
This type of fluctuation corresponds to case B. It can be used to actively control $\nu_{ceo}$, 
whereas cavity length control is suited to stabilize the optical carrier frequency.
It follows from the elastic tape picture, that technical noise contributions from the frequency comb are 
completely known if the phase angles of two distinct comb lines are monitored as a function of time. 
Two candidates which naturally come to mind are $\nu_{ceo}$ and an arbitrary line at $\nu_m$ within 
the span of the comb. The latter has to be measured with respect to a stable  optical reference frequency 
at $\nu_x$, i. e. by measurement of the beat-note frequency
$\Delta_x(t)=\nu_x-\nu_m(t)$. We will assume that the signal-to-noise ratios of 
both the $\Delta_x$ and $\nu_{ceo}$ beat-notes are sufficient to yield an rms-phase error 
of $<0.1$~radians within a bandwidth which is sufficient to track technical fluctuations of 
$\Delta_x$ and $\nu_{ceo}$. Cycle-slipping can be excluded under such conditions. 
Then, both $\varphi_m$ and $\varphi_{ceo}$ are unambiguously known and so are the instantaneous 
phase angles of all other comb lines, according to Eqn.~(\ref{phim}). 
This feed-forward technique for phase angles is the key element of the transfer concept. 
In the following, we present two applications of this method.

\vspace{3ex}
\noindent {\bf 3. Linking optical and microwave frequencies}
\vspace{1ex}

The first application deals with the frequency modulation (FM) noise analysis of a 
microwave signal if an optical signal with frequency $\nu_x$ is available which shows 
superior FM noise properties. 
The scheme is shown in  Fig.~1. Here, three input signals are detected by photo diodes:

\begin{enumerate}
\item[i.] the beat note $\Delta_x$ between the external signal at $\nu_x$ and the nearest comb line,
\item[ii.] the pulse repetition frequency $f_{rep}$, and
\item[iii.] the carrier-envelope-offset frequency $\nu_{ceo}$.
\end{enumerate}

\begin{figure}[ht]
  \centerline{\includegraphics[width=8cm]{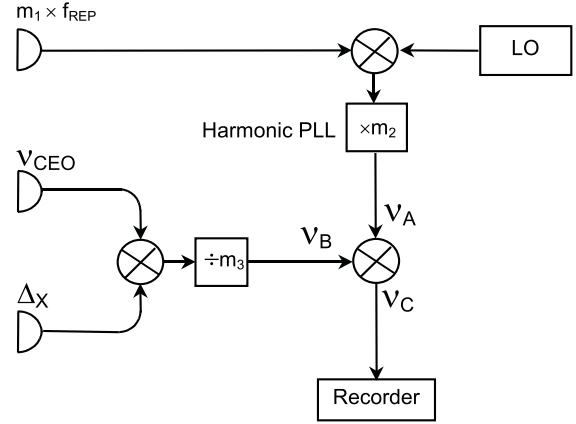}}\smallskip
  \caption{Linking microwave and optical frequencies: signal processing scheme.}
\end{figure}

The detection of $\nu_{ceo}$ requires one or more additional nonlinear processes [\ref{tell99}]. 
The simplest case is applicable if the comb covers a frequency ratio of more than a factor of two, 
i.e. one octave. 
Then, the comb lines at the low frequency comb wing are frequency doubled to yield a 
beat note of their second harmonics with the lines at the high frequency comb wing. 
All comb lines contain one $\nu_{ceo}$ according to Eqn.~(\ref{num}) but the second 
harmonics contain two $\nu_{ceo}$. 
Thus, the beat note oscillates at $\nu_{ceo}$ providing the desired input signal. 
If all three signals are available, one has to select an order number $m_x$ 
which can be separated into three factors, 
each of the order of $10^2\!: \; m_x = m_1\times m_2 \times m_3$. 
This can be accomplished choosing a proper value of $f_{rep}$. Then, the frequency
$f_{LO}$ of the microwave local oscillator (LO) is mixed with a harmonic $m_1$ of the
pulse repetition frequency as detected by a photo diode to yield an rf-signal at frequency
$f_{LO}-m_1f_{rep}$. This frequency is further multiplied by a factor of $m_2$ with the 
help of a harmonic PLL yielding

\begin{equation}
\nu_A  = m_2f_{LO} - m_1m_2f_{rep} \;.
\label{nua}
\end{equation}

In a second channel, the sum frequency of $\nu_{ceo}$ and $\Delta_x$ 
is divided by $m_3$. This leads to a signal at frequency 

\begin{equation}
\nu_B  = \frac{1}{m_3} (\nu_{ceo} + \Delta_x ) \;.
\label{nub}
\end{equation}

Subtracting both frequencies with the help of a mixer, as shown in Fig.1, results in

\begin{equation}
\nu_C = \nu_A - \nu_B = m_2f_{LO} - \left( m_1m_2f_{rep} + \frac{\nu_{ceo} + \Delta_x}{m_3} \right) \;.
\label{nuc}
\end{equation}

However, the elastic tape formula (1) predicts for the expression in the bracket 

\begin{equation}
m_1m_2f_{rep} + \frac{\nu_{ceo} + \Delta_x}{m_3} = \frac{1}{m_3}\,\nu_x  \;,
\label{nux}
\end{equation}

\noindent and thus

\begin{equation}
\nu_C = m_2f_{LO} - \frac{\nu_x}{m_3} \;.
\label{nucend}
\end{equation}

Note that  the signal at $\nu_C$ is independent of the noise properties of the 
KLM laser, i.e. the laser acts as a true transfer oscillator, bridging a frequency 
ratio of $f_{LO}/\nu_x$. 
Since the phase angles of  all signals are processed according to 
Eqn.~(\ref{phim}), $\nu_C$ can be considered as the frequency of the beat-note 
between the $m_2$th harmonic of $f_{LO}$ and the $m_3$th sub-harmonic of $\nu_x$.

\vspace{3ex}
\noindent {\bf 4. Linking two optical frequencies}
\vspace{1ex}

The second application of the transfer concept deals with the FM-noise measurement 
of an optical signal with frequency $\nu_y$ if an optical reference is available which has superior 
noise properties but oscillates at a very different frequency $\nu_z$, as shown in Fig.~2. 
\begin{figure}[ht]
  \centerline{\includegraphics[width=6.5cm]{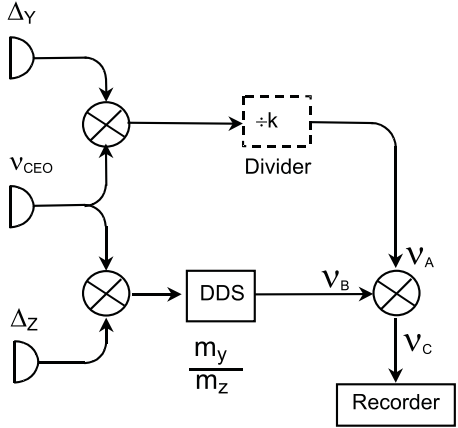}}\smallskip
  \caption{ Linking two optical frequencies: signal processing scheme. The frequency divider is optional.}
\end{figure}
As before, we apply the elastic tape picture, which leads to

\begin{equation}
\nu_y = m_y f_{rep} + \nu_{ceo} + \Delta_y	
\label{nuyopt}
\end{equation}	
 
\noindent and

\begin{equation}
\nu_z = m_z f_{rep} + \nu_{ceo} + \Delta_z \;.
\label{nuzopt}
\end{equation}	

The beat signal at $\Delta_y$ is mixed with the $\nu_{ceo}$-beat signal which leads to 

\begin{equation}
\nu_A = \nu_{ceo} + \Delta_y	
\label{nuaopt}
\end{equation}	

\noindent if we neglect the optional frequency divider for the moment. 
The sum frequency of $\Delta_z$ and $\nu_{ceo}$, on the other hand, 
is processed with a direct-digital-synthesis IC (DDS). 
Such a device is capable of generating an output signal from an input clock signal 
with a frequency ratio given by a long digital tuning word while the input-to-output 
frequency agility is preserved. 
The latency time of these devices, i.e. the time required to set a new tuning word, 
is not a problem in our case, since the tuning word is chosen once and then held fixed. 
The DDS numerically approximates the ratio of two integers $m_y/m_z$ by $j/2^n$, 
where $j$ is an integer and $n$ the bit-length of the tuning word. 
As discussed below, the resulting error is negligible for $n=32$~bit or 48~bit in our case, 
where the integers $m_y$ and $m_z$ are of the order of $10^6$.
As a result of this signal processing, one obtains 

\begin{equation}
\nu_B = \frac{m_y}{m_z} (\nu_{ceo} + \Delta_z) \;.	
\label{nubopt}
\end{equation}	

\noindent Generating the difference frequency between both signals as before 
and using Eqns.~(\ref{nuyopt}) and (\ref{nuzopt}), we find

\begin{equation}
\nu_C = \nu_A - \nu_B = \nu_y - \frac{m_y}{m_z}\nu_z \,,	
\label{nucopt}
\end{equation}	

\noindent which is again independent of the properties of the KLM laser. 
Since the phase angles of all signals are processed according to 
Eqn.~(\ref{phim}) as above, $\nu_C$ can be considered as the frequency of the beat-note 
between $\nu_y$ and $\nu_z$ as {\it projected} to the vicinity of $\nu_y$.

\vspace{3ex}
\noindent {\bf 5. Experimental}
\vspace{1ex}

The setup of our Kerr-lens mode-locked Ti:Sapphire-laser is similar to [\ref{sutt99}], employing 
a combination of prisms and double chirped mirrors for compensation of group velocity dispersion. 
A similar, external prism pair is used for pulse re-compression. 
The pulse duration is $<20$~fs (FWHM) while the output spectrum typically spans 70~THz (FWHM), 
centred at about 790~nm. The pulse repetition frequency is about 100~MHz. 
Approximately 30~mW of the laser output is coupled into a 10~cm long piece of air-silica 
microstructure (MS) fiber with a core diameter of 1.7~$\mu$m and a zero-GVD wavelength 
of 780~nm [\ref{rank00}]. 
The output spectrum of the fiber extends from about 500~nm to about 1100~nm. 
The carrier-envelope-offset frequency $\nu_{ceo}$ is measured by second-harmonic-generation of the 
comb's infrared portion around 1070~nm in a nonlinear-optical crystal (LBO). 
The beat-note between the resulting green SHG signal and the green output of the MS fiber is 
detected by a photo multiplier (PM) after spectral and spatial filtering both fields with a single mode 
fiber and a 600~l/mm grating, respectively. 

As frequency references in the optical range, we use three different signals,
\begin{enumerate}
\item[i)] the sub-harmonic at 344~179~449~MHz (871~nm) of the output of a single Yb$^+$-ion 
            frequency standard [\ref{tamm00}],
\item[ii)] the output of a Nd:Yag laser at 281~606~335~MHz (1064~nm) which is frequency-stabilized 
via saturated absorption of its second harmonic in I$_2$ vapour, and 
\item[iii)] the output of a dye-laser that is frequency-stabilized to the Ca intercombination line at 
455~986~240~MHz (657~nm) [\ref{rieh99}]. 
\end{enumerate}

All three frequencies have been previously measured with respect to a Cs atomic 
clock [\ref{udem01}], [\ref{sten01b}], [\ref{nevs01}], [\ref{sten01}], [\ref{schn96}]. 
However, these absolute values are not important for the purpose of this paper, 
which aims to demonstrate a novel measurement and synthesis principle.
The sources of the signals at 344~THz, 281~THz and 455~THz will be referred to as 
Yb-, Iodine- an Ca-standard in the following (index Yb, Iod and Ca).

The Yb-standard was employed both for the micro\-wave-to-optical and the optical-to-optical link. 
The pulse repetition frequency of the laser was set to a value close to 100~MHz which resulted
in a mode number $m_{Yb}$~=~3~441~024 for the mode closest to $\nu_{Yb}$.
As discussed above, this number must be divisible by 3 factors of the order of 
$10^2$ such as $m_{Yb}=m_1\times m_2\times m_3=103\times 96\times 348=3~441~024$. 
The pulse repetition frequency was measured with a fast InGaAs photo diode (PD) 
at ~10.3~GHz, i.e. at a harmonic order of $m_1=103$. 
For the sake of dynamic range of the PD we restricted the number of detected harmonic 
orders to a few using optical pre-filtering in a Fabry-Perot interferometer with a free spectral range 
of about 10.3~GHz (10~mm thick fused silica etalon). 
The output signal of the microwave PD was down-converted with the help of a double-balanced mixer 
and a microwave synthesizer (LO) controlled by a 100~MHz standard frequency 
from a H-maser. The LO frequency at $f_{LO}\approx 10.3$~GHz was tuned to yield a 
down-converted signal of about 500~kHz. 
This frequency was multiplied by a factor of $m_2=96$ with the help of a harmonic PLL. 
Hence, its output signal at $f_1\approx 48$~MHz carried the frequency noise of the 
$(m_1 m_2)$th harmonic of the pulse repetition frequency and that of the 
H-maser as multiplied to a virtual frequency of $m_2 f_{LO} \approx 989$~GHz. 
As described above, the frequency noise of the pulse repetition frequency can independently be 
deduced from $\nu_{ceo}$ and $\Delta_{Yb}$. 
Both signals were pre-filtered with PLLs with a servo bandwidth of $> 1$~MHz. 
As a result of choice of signs, the sum frequency of both ($\approx$~70~MHz) carried the 
desired information. 
This signal was frequency divided by $m_3=348$ leading to a signal at $\nu_B \approx 200$~kHz 
which carried the noise of the $(m_{Yb}/m_3)$th harmonic of the pulse repetition frequency and 
that of the Yb standard as divided to a virtual frequency of $\nu_{Yb}/m_3 \approx 989$~THz. 

Since $m_1m_2 = m_{Yb}/m_3$, the noise of the pulse repetition frequency was exactly 
the same in both paths and cancelled out if the frequency difference between
$\nu_A$ and $\nu_B$ was generated with the help of the last mixer, as shown in Fig.~1, 
thus realising the transfer principle. As mentioned above, the output of this mixer at
$\nu_C = \nu_B - \nu_B \approx 47.8$~MHz can be considered as the beat-note between the 
H-maser signal as multiplied to 989~GHz and the output of the Yb standard, 
as divided by 348 to 989~GHz. The signal at $\nu_C$ was down-converted to about 40~Hz 
with the help of an rf-synthesizer and a mixer, analog-to-digital converted, digitally recorded and 
subsequently Fourier-transformed. 

\begin{figure}[ht]
  \centerline{\includegraphics[width=10cm]{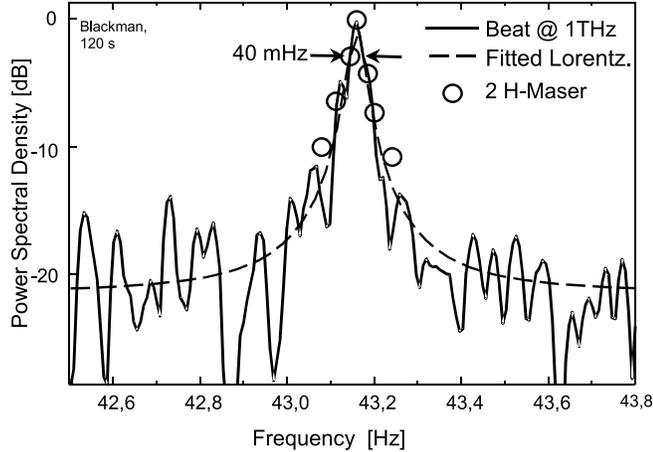}}\smallskip
  \caption{Beat-note spectrum between a harmonic of the H-maser signal and a sub-harmonic of 
the output of the optical Yb$^+$ frequency standard taken at a frequency of 989~GHz. 
The circles represent the beat-note of the signals of two H-maser as extrapolated to the same frequency.}
\end{figure}
A spectrum of such a 120~s long record is shown in 
Fig.~3. One finds a 40 mHz wide carrier on top of a much broader line which can not be seen in Fig.~3. 
Note that this line with a sub-Hertz width was generated from two signals whose frequencies 
fluctuated by many kHz which demonstrates the effect of the transfer concept. 
The control bandwidth of the harmonic PLL was set to a few tens of kHz. 
This resulted, together with the second order characteristic of its loop filter, to a loop gain of $>100$~dB 
in the Fourier frequency range shown in Fig.~3. 
The loop gains were even higher ($>\!130$~dB) for the PLLs which phase-tracked the 
signals at $\nu_{ceo}$ and $\Delta_{Yb}$. 
Thus, the sum of the residual servo errors, i.e. the total error of the frequency transfer process, 
was negligible and Fig.~3 can be considered to represent the true line shape of the 
H-maser harmonic at 989~GHz since the frequency noise of the Yb signal 
was also negligible as shown below.

The shape of the spectrum, a narrow line on top of a broad pedestal, was expected from the 
phase-noise specifications of the H-maser. 
This broad pedestal becomes dominant and submerges the narrow carrier, if the H-maser 
signal is further frequency multiplied, e.g. to the optical range. 
Thus, the short-term instability of a frequency ratio measurement between a H-maser and a quiet 
optical frequency standard, such as our Yb- or Iodine standards, is expected to be limited by the H-maser noise.

For comparison, the open circles in Fig.~3 show the corresponding beat note spectrum of the 
outputs of two H-maser as multiplied to a frequency of 989~GHz. 
It was calculated from a 1000~s record of readings of the timing jitter between 10~MHz standard 
frequencies generated by both masers. 
The close agreement with the Yb/H-beat at 989~GHz demonstrates that the frequency multiplication 
process by a factor of $10^6$ performed by the KLM laser does not deteriorate the 
H-maser noise properties at Fourier frequencies below 0.1~Hz.

In the following, we describe a second  application of the transfer concept, the measurement 
of an optical frequency ratio. Here, an additional photo diode was used to detect the beat-note 
at $\Delta_{Iod}$ ($\approx 44$~MHz) between the comb line with order number
$m_{Iod} = 2~815~433$ and the output of the Iodine-standard. 
As mentioned above, the sum frequency $\Delta_{Iod}+\nu_{ceo}$ was processed by the 
DDS after pre-filtering both signals with fast PLL tracking oscillators (bandwidth $>\!1$~MHz) . 
Since the required multiplication factor $m_{Yb}/m_{Iod}$ was of the order of one, which can 
not be generated by a DDS chip according to the sampling theory, 
we used an additional division factor of $k_{Iod}=8$ in both signal paths in Fig.~2. 
The divider in the Yb-signal path was a conventional TTL divider while in the other path the 
DDS tuning word was corrected for this value. 
The DDS (AD9851) was programmed within the resolution of its 32~bit tuning word to generate 
an output frequency $f_{out} = 0.152775079245 \times f_{in}$. 
This was 8.4~mHz lower than the required value $(m_{Yb}/m_{Iod})/8 = 0.152775079357\cdots $.
However, multiplication of this amount by 8 and division by 344~THz leads to an relative error of
$ 2 \cdot 10^{-16}$ which is negligible compared to other uncertainties. 
However, for future applications which might require higher precision, 
this error can be reduced by at least 4 orders of magnitude with a proper correction or with a 
DDS with 48~bit tuning word.

The output of the last mixer in Fig.~2, which was equivalent to the beat signal between the 
Iodine- and Yb- signals at a virtual frequency of 344/8~THz~=$\!$~43~THz, was down-converted to the kHz range, 
analog-to-digital converted and digitally recorded. 
The Fourier transform of a typical 30~s record is shown in Fig~4. One finds a $11.5$~Hz wide line 
on top of a pedestal of white additive noise. 
As in Fig.~3, the PLL's bandwidths of $>\!1$~MHz ensured a negligible residual servo error. 
Thus, Fig.~4 shows the true power spectrum of a non-integer sub-harmonic of the Iodine signal 
at 43~THz since the contribution of the Yb signal to the spectrum was not significant, as discussed below.

\begin{figure}[ht]
  \centerline{\includegraphics[width=9cm]{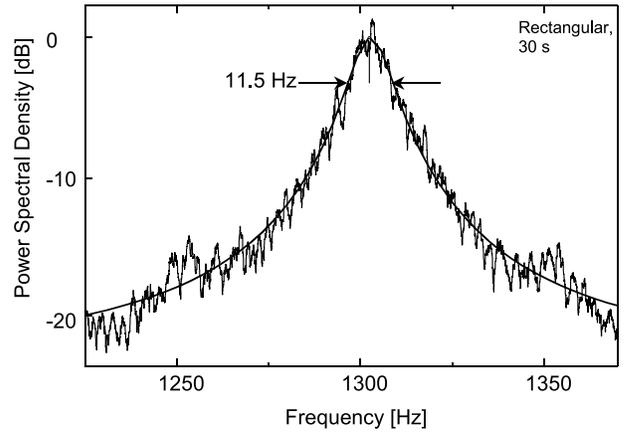}}\smallskip
  \caption{FFT-spectrum of the beat note between sub-harmonics of the outputs of an
$I_2$ and an Yb$^+$ frequency standard taken at a frequency of 43~THz.}
\end{figure}

Finally, we describe a third frequency ratio measurement employing the Yb- and Ca-standards. 
The mode order of comb line nearest to the Ca frequency was $m_{Ca}=4~558~841$ 
while $m_{Yb}=3~441~024$ during all experiments. 
Here, the Ca frequency was down-converted as opposed to the experiment described above, 
in which the Iodine-frequency of 281~THz was up-con\-verted to $m_{Yb}=344$~THz. 
For this reason, the pre-division factor could be reduced to $k_{Ca}=4$. 
Consequently, the output of the last mixer in Fig.~2 corresponded to the beat-note between the 
Yb and Ca signals at a frequency of 344/4~THz~=~86~THz. 
\begin{figure}[ht]
  \centerline{\includegraphics[width=9cm]{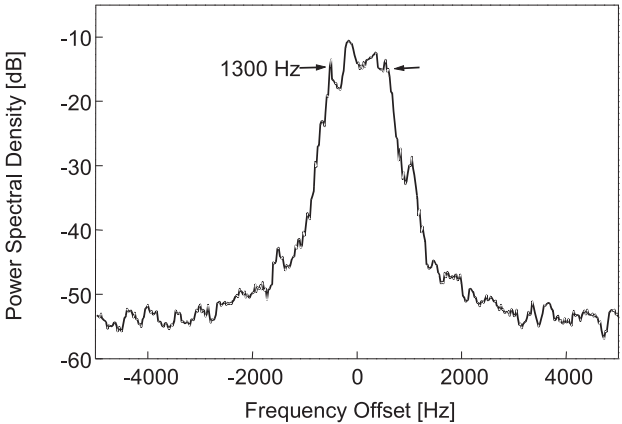}}\smallskip
  \caption{Spectrum of the beat-note between the outputs of an 
  Yb$^+$ and Ca optical frequency standard taken with a conventional rf-spectrum analyzer. 
Resolution bandwidth 300~Hz, center frequency 6.234~MHz.}
\end{figure}
The spectrum in Fig.~5, however, 
shows the 4th harmonic of this signal as generated by a frequency-quadrupling PLL. 
Thus, it represent the beat-note at the Yb-frequency which means that the frequency transfer 
was carried out over a frequency gap of ($455- 344$)~THz~=~111~THz. 
The width of 1300~Hz found in Fig.~5 was mainly due to low-frequency jitter 
of the Ca frequency while the line width of the Yb-signal was much smaller. 
This was proved by resolving a 30~Hz wide resonance line at its second harmonic [\ref{sten01b}].

As mentioned above, the frequency noise of the beat-note between the Yb- and Iodine-signals 
is substantially smaller than that of the H-Maser, at least at Fourier frequencies $f>0.1$~Hz. 
Thus, a frequency ratio measurement of these two optical signals can be carried out with 
smaller instabilites than that of an Yb/H-maser frequency ratio measurement, at least for short 
averaging times. This is demonstrated in Fig.~6. 
\begin{figure}[ht]
  \centerline{\includegraphics[width=10cm]{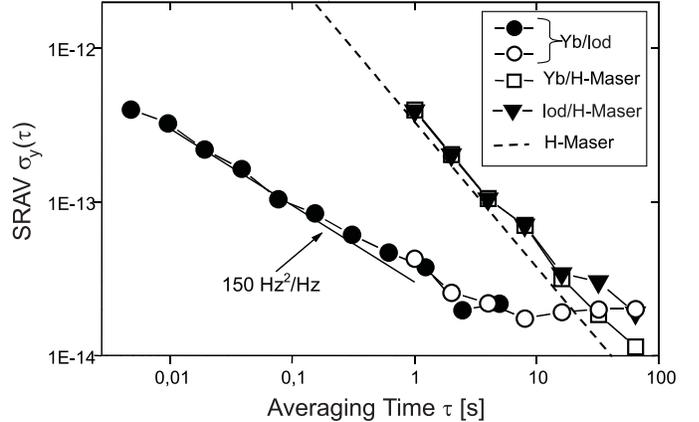}}\smallskip
  \caption{Allan standard deviation of three frequency ratio measurements: 
Yb-standard/H-maser, Iodine-standard/H-maser and Yb-/Iodine-standard. 
Note that the instability of the optical frequency ratio measurement is substantially smaller than that 
of the H-maser for averaging times below 10~s.}
\end{figure}
The dashed curve shows the sqare root of the Allan 
variance $\sigma_y (\tau)$ (SRAV) of our H-maser. The data of the frequency measurements of the 
Iodine and the Yb-standard with respect to this H-maser are shown as triangles and squares, 
respectively. The SRAV of the Yb-laser qualitatively follows that of the H-maser, which 
limits the measurement  for averaging times from 1 to 100~s. 
The SRAV of the Iodine -signal is larger than that of the H-maser for $\tau >20$~s. 
In fact, frequency comparisons of two of such Iodine standards showed a SRAV of about 
$2 \cdot 10^{-14}$ in this range and thus larger than of the H-maser. 
The SRAV of a frequency ratio measurement of two optical standards, on the other hand, 
is not limited by H-maser noise, as shown by circles in Fig.~6. 
The data depicted by solid circles have been derived from the Yb/Iodine beat signal as obtained 
from the last mixer in Fig.~2 whereas the open circles have been calculated from
$\nu_{ceo}, \Delta_{Yb}$, and $\Delta_{Iod}$ as counted by totalizing counters (1~s averages). 
Both data sets reasonably agree in the overlapping range between 1 and 8~s. 
The SRAV values of this frequency ratio measurement are substantially smaller than the 
H-maser frequency instability for $\tau <10$~s. The $1/\sqrt{\tau}$-dependency of this SRAV function 
indicates a white frequency noise level of $S_\nu \approx 150$~Hz$^2$/Hz at 344~THz which would 
result in a spectral line width of $\delta\nu =\pi S_\nu\approx 500$~Hz (FWHM) 
in good agreement with the direct beat-note measurement as shown in Fig.~4. 
From $\delta\nu \approx 11.5$~Hz at (344/8=43)~THz one calculates
$\delta\nu = 8^2\cdot 11.5 {\rm Hz} \approx 730$~Hz  at 344~THz 
under the assumption of white frequency noise. 
This indicates, that the Yb/Iodine-frequency ratio measurement in Fig.~6 was limited by noise of the 
Iodine-standard whereas contributions from the Yb-signal were negligible due to its narrow line 
width of $< 30$~Hz.

\vspace{3ex}
\noindent {\bf 6. Conclusion}
\vspace{1ex}

We have demonstrated a novel concept for frequency measurement and synthesis 
which is capable of phase-coherently linking signals from very different spectral regions in the optical and 
microwave ranges without introducing additional noise. 
We have carried out frequency ratio measurements between optical frequencies with short-term 
instabilities superior to that of a microwave reference. Since the measurement uncertainties were clearly 
limited by the noise properties of the frequency standards, one may expect an even lower limitation due 
to noise contributions of the KLM laser if better optical frequency standards become available. 
The transfer concept opens up new perspectives for future ultra-high precision applications, e.g. 
measurement of time variations of fundamental constants as soon as appropriate optical frequency 
standards are available.

We gratefully acknowledge financial support from the Deutsche Forschungsgemeinschaft through 
SFB407 and contributions by Andreas Bauch, Tomas Binnewies, Nils Haverkamp, Ursula Keller, 
Harald Schnatz, G\"unter Steinmeyer, Christian Tamm, and Guido Wilpers in different stages of the experiments. 
We are also indebted to Robert Windeler of Lucent Technologies for providing us with the microstructure fiber.

\vspace{3ex}
\noindent {\bf 6. References}

\renewcommand{\labelenumi}{[\arabic{enumi}]}
\begin{enumerate}

\item\label{udem99}
T. Udem, J. Reichert, R. Holzwarth, and T. W. H\"ansch; Opt. Lett. {\bf 24} 881 (1999).

\item\label{udem01}
T. Udem, S. A. Diddams, K. R. Vogel, C. W. Oates, E. A. Curtis, W. D. Lee, W. M. Itano, 
R. E. Drullinger, J. C. Bergquist, and L. Hollberg; Phys. Rev. Lett. {\bf 86}  4996  (2001).

\item\label{sten01b}
J. Stenger, Chr. Tamm, N. Haverkamp, S. Weyers, and H. R. Telle; Opt. Lett., in press (2001).

\item\label{nier00}
M. Niering, R. Holzwarth, J. Reichert, P. Pokasov, Th. Udem, M. Weitz, T. W. H\"nsch, 
P. Lemonde, G. Santarelli, M. Abgrall, P. Laurant, C. Salomon, and A. Clairon; Phys. Rev. Lett. {\bf 84}  5496  (2000).

\item\label{jone00}
D. J. Jones, S. A. Diddams, J. K. Ranka, A. Stentz, R. S. Windeler, J. L. Hall, and S. T. Cundiff; 
Science {\bf288}, 635 (2000).

\item\label{kram92}
G. Kramer, B. Lipphardt, and C. O. Weiss, Proc. 1992 Frequ. Contr. Symp., {\bf 39} (1992), 
IEEE Cat. No. 92CH3083-3.

\item\label{reic99}
J. Reichert, R. Holzwarth, Th. Udem, and T.W. H\"ansch; Opt. Comm. {\bf172} 59 (1999).

\item\label{tell99}
H.R. Telle, G. Steinmeyer, A.E. Dunlop, J. Stenger, D. H. Sutter, and U. Keller; Appl. Phys. {\bf B69} 327 (1999).

\item\label{sutt99}
D. H. Sutter, G. Steinmeyer, L. Gallmann, N. Matuschek, F. Morier-Genoud, U. Keller, V. Scheuer, 
G. Angelow, and T. Tschudi; Opt. Lett. {\bf 24} 631 (1999).

\item\label{rank00}
J. K. Ranka, R. S. Windeler, and A. J. Stentz; Opt. Lett. {\bf 25} 25 (2000).

\item\label{tamm00}
Chr. Tamm, D. Engelke, V. B\"uhner; Phys. Rev. {\bf A61} 053405 (2000).

\item\label{rieh99}
F. Riehle, H. Schnatz, B. Lipphardt, G. Zinner, T. Trebst, and J. Helmcke;
IEEE Trans. Instr. Meas. {\bf IM48} 613 (1999).

\item\label{nevs01}
A.Y. Nevsky , R. Holzwarth, J. Reichert, Th. Udem, T.W. H\"ansch, J. von Zanthier, H. Walther, 
H. Schnatz, F. Riehle, P.V. Pokasov, M.N. Skvort\-sov, and S.N. Bagayev; Opt. Comn. {\bf 192} 263 (2001).

\item\label{sten01}
J. Stenger, T. Binnewies, G. Wilpers, F. Riehle, H. R. Telle, J. K. Ranka, R. S. Windeler, A. J. Stentz;
Phys. Rev. A. {\bf 63} 021802(R) (2001).

\item\label{schn96}
H. Schnatz, B. Lipphardt, J. Helmcke, F. Riehle, and G. Zinner; Phys. Rev. Lett. {\bf 76} 18 (1996).

\end{enumerate}

\end{document}